\input{aipcheck}

\documentclass[
  ,draft            % use draft while you are working on the paper
  ]
  {aipproc}

\layoutstyle{6x9}

\begin{document}

\title{The Pb Radius Experiment (PREX)}


\author{Juliette M. Mammei}{
  address={University of Massachusetts, Amherst and Jefferson Lab, Newport News, VA 23606}
}

\begin{abstract}
We report the first measurement of the parity-violating asymmetry A$_{PV}$ in the elastic scattering of polarized electrons from $^{208}$Pb from the Lead Radius Experiment PREX which ran in Hall A at the Thomas Jefferson National Accelerator Facility (JLab). A$_{PV}$ is sensitive to the radius of the neutron distribution R$_n$. The Z boson that mediates the weak neutral interaction couples mainly to neutrons and provides a clean, model-independent measurement of the RMS radius R$_n$ of the neutron distribution in the nucleus and is a fundamental test of nuclear structure theory.  The result, A$_{PV}$ = 0.656 $\pm$ 0.060(stat) $\pm$ 0.014(syst) ppm, corresponds to a difference between the radii of the neutron and proton distributions R$_n$ - R$_p$ = 0.33 $^{+0.16} _{-0.18}$ fm and provides the first electroweak observation of the neutron skin which is expected in a heavy, neutron-rich nucleus.
\end{abstract}

\maketitle

%%%%%%%%%%%%%%%%%%%%%%%%%%%%%%%%%%%%%%%%%%%%
%% MAINMATTER
%%%%%%%%%%%%%%%%%%%%%%%%%%%%%%%%%%%%%%%%%%%%

\section{Overview}
Combining known nuclear charge densities, which have been determined accurately with electron scattering \cite{Frois:1977}, with a measurement of the weak charge density allows us to determine the neutron skin of a heavy nucleus.  Measurement of the parity-violating asymmetry in electron scattering provides a model-independent probe of neutron densities that is free from most strong-interaction uncertainties because the weak charge of the neutron is much larger than that of the proton \cite{DonnelyDubachSick:1989}.  In the Born approximation, the asymmetry for longitudinally polarized electrons scattered elastically from an unpolarized nucleus is proportional to the weak form factor $F_W(Q^2)$.  The Born approximation is not valid for a heavy nucleus, and Coulomb-distortion effects must be taken into effect \cite{Horowitz:1998}.  The asymmetry is \begin{equation}
A_{PV}= \frac{\sigma_R-\sigma_L}{\sigma_R+\sigma_L}\approx\frac{G_FQ^2}{4\pi\alpha\sqrt{2}}\frac{F_W(Q^2)}{F_{ch}(Q^2)}\label{pvasym}
\end{equation} where $\sigma_{R(L)}$ is the differential cross section for elastic scattering of right- (R) and left- (L) handed longitudinally polarized electrons, $G_F$ is the Fermi constant, $\alpha$ the fine structure constant, and $F_{ch}(Q^2)$ is the Fourier transform of the known charge density distribution.  Likewise, the weak form factor, $F_W(Q^2)$, is the Fourier transform of the weak charge density; the first derivative is related to the RMS of the distribution, so measurement at one $Q^2$ is sufficient to constrain the neutron skin (the difference in the RMS of the neutron and proton distributions).  Other details relevant for a parity-violation measurement of neutron densities have been discussed in \cite{HorowitzPollockSouderMichaels:2001}.  

The weak form factor, when combined with measurements of the charge form factor, can then be used to extract the difference between the radii of the neutron and proton distributions, or the ``neutron skin''.  There is a strong correlation between the neutron radius, $R_n$ and the equation of state of neutron matter \cite{Brown:2000}.  Therefore, measuring the neutron radius has important implications for nuclear astrophysics, particularly for the structure of neutron stars.

\section{The PREX Experiment}

The measurement was performed in Hall A of Thomas Jefferson National Accelerator Facility (JLab).  It made use of the two standard Hall A high resolution spectrometers and a septum magnet to achieve the $\theta_{lab}\sim$5$^{\circ}$ bend angle in each arm.  The beam consisted of 1.605 GeV longitudinally polarized electrons, with a current of 40 to 70 $\mu$A.  The polarized beam is produced from circularly polarized laser light incident on a strained GaAs photocathode \cite{source1, source2}.  The polarization of the beam was pseudo-randomly changed at 120 Hz, resulting in 8.33 ms ``helicity windows'' with either positive or negative helicity laser light (right- or left- handed electrons).  The target consisted of 0.55 mm of an isotopically pure $^{208}$Pb foil sandwiched between two 150 $\mu$m diamond foils (for heat transfer).  The beam was rastered over the face of the target in a 4$\times$4 mm square in order to prevent target melting.  

The detectors consisted of thick and thin quartz bars with air light guides to transport the Cerenkov light to photomultiplier tubes.  The resolution of the spectrometers ensured that only elastic events and a negligible fraction of inelastic events from the first excited state of $^{208}$Pb (2.6 MeV) made it onto the detectors.  The signals were integrated over the helicity windows and the asymmetry was calculated from quartets (either + - - + or - + + -).  The beam quality was monitored so that fluctuations in beam intensity, energy or position were negligible compared to those from statistical fluctuations.  The weak charge form factor is obtained from the measurement of the asymmetry \cite{prexprc}.  The asymmetry, A$_{PV}$ = 0.656 $\pm$ 0.060(stat) $\pm$ 0.014(syst) ppm, corresponds to a neutron skin of R$_n$ - R$_p$ = 0.33 $^{+0.16} _{-0.18}$ fm \cite{prexprl}.

\section{Future Plans}

During the PREX-I run the systematic error goals of the experiment were achieved, although the statistical error bar was larger than originally proposed.  The run was limited by multiple repairs to a vacuum seal and electronics which were damaged by radiation.  The experiment is being redesigned to reduce the effect of radiation in the hall, in particular, neutron backgrounds which were the primary cause of damage to the electronics.  In addition the o-ring, which was used due to other engineering constraints, will be replaced with a metal seal and newer, more radiation-hard electronics will be installed. 

\begin{figure}
  \includegraphics[height=.3\textheight]{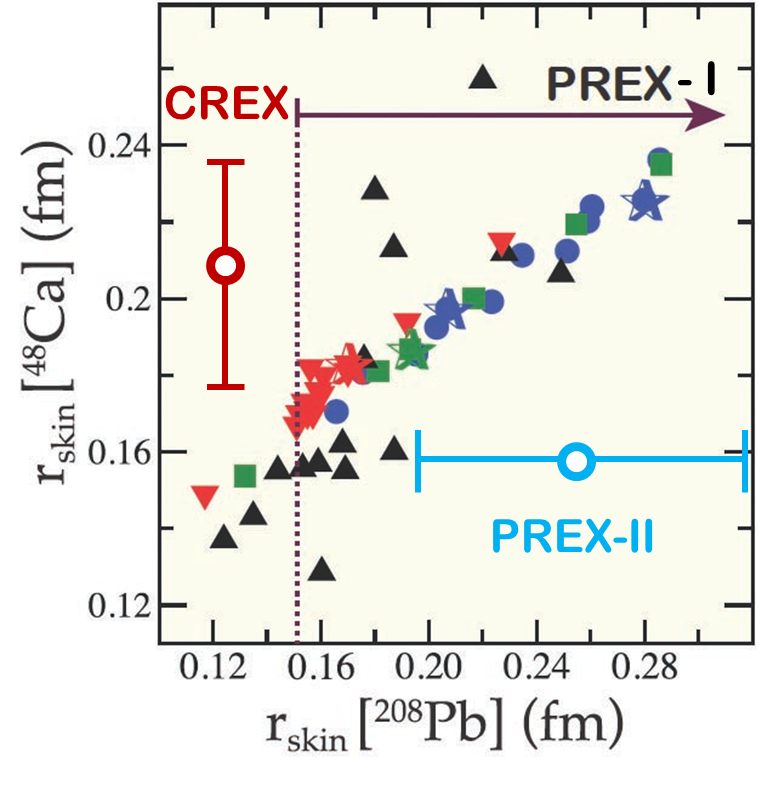}
  \caption{Plot of the model predictions of the neutron skin of $^{48}$Ca vs. that of $^{208}$Pb, with lower limit from PREX I (the central value is off the plot to the right).  The estimated errors on the future PREX II and CREX are also shown (open circles).}\label{fig:prexcrex}
\end{figure}

PREX-II has been approved to achieve the originally proposed error bar.  The CREX experiment, to measure the neutron skin of $^{48}$Ca, has also been conditionally approved.  Alone, the measurement of the neutron skin in the lighter $^{48}$Ca nucleus will make it possible to test the predictions of detailed microscopic models of nuclear structure involving three neutron forces which are not possible with heavier nuclei.  It will increase the precision of the neutron skin measurement, because the neutron skin of the two nuclei is expected to be highly correlated, though the larger $^{208}$Pb nucleus more closely approximates infinite nuclear matter.  Figure \ref{fig:prexcrex} shows the predictions of the neutron skin of $^{48}$Ca vs. that of $^{208}$Pb for a number of state-of-the-art nuclear structure models \cite{Piekarewicz:2012}.  In combination, the results from the PREX and CREX measurements will challenge the assumptions of these models by testing the dependence on the atomic mass number A.

%%%%%%%%%%%%%%%%%%%%%%%%%%%%%%%%%%%%%%%%%%%%%%%%
%% BACKMATTER
%%%%%%%%%%%%%%%%%%%%%%%%%%%%%%%%%%%%%%%%%%%%%%%%

\begin{theacknowledgments}
We wish to thank the entire staff of JLab for their efforts to develop and maintain the polarized beam and the experimental apparatus. This work was supported by the U.S. Department of Energy, the National Science Foundation, and from the French CNRS/IN2P3 and ANR. Jefferson Science Associates, LLC, operates Jefferson Lab for the U.S. DOE under U.S. DOE Contract No. DE-AC05-060R23177.
\end{theacknowledgments}

\bibliographystyle{aipproc}

\bibliography{021_Mammei}

\end{document}